\documentclass[aps,prl,twocolumn,superscriptaddress]{revtex4-2}

\usepackage{graphicx}
\usepackage{amsmath,amssymb,latexsym}
\usepackage{color}
\usepackage{xcolor}
\usepackage{ulem}
\usepackage{array}
\usepackage{hyperref}
\hypersetup{
  colorlinks,
  citecolor=blue,
  linkcolor=red}
\newcommand{\be}{\begin{equation}}
\newcommand{\ee}{\end{equation}}
\newcommand{\beq}{\begin{eqnarray}}
\newcommand{\eeq}{\end{eqnarray}}
\newcommand{\bee}{\begin{equation*}}
\newcommand{\eee}{\end{equation*}}
\newcommand{\beqq}{\begin{eqnarray*}}
\newcommand{\eeqq}{\end{eqnarray*}}
\newcommand{\w}{\omega}
\newcommand{\W}{\Omega}
\newcommand{\g}{\gamma}

\newcommand{\De}{\Delta}
\newcommand{\de}{\delta}

\newcommand{\s}{\sigma}

\newcommand{\ket}[1]{\left| #1 \right\rangle}
\newcommand{\bra}[1]{\left\langle #1 \right|}

\newcommand{\sbrac}[1]{\langle #1 \rangle}

\newcommand{\TC}{{\cal T}_{\cal C}}
\newcommand{\SC}{{\cal S}_{\cal C}}
\newcommand{\Se}{{\cal S}^{\rm eff}_{\cal C}}

\newcommand{\squ}[1]{[ #1 ]}

\newcommand{\ih}{-\frac{i}{\hbar}}

\newcommand{\hide}[1]{}

\newcommand{\eq}[1]{Eq.\,(\ref{#1})}
\newcommand{\eqs}[1]{Eqs.\,(\ref{#1})}

\newcommand{\fig}[1]{Fig.\,\ref{#1}}

\graphicspath{{Figures/}}

\begin{document}


\title{Collective Lamb Shift and Modified Linewidth of An Interacting Atomic Gas}


\author{Hanzhen Ma}
\affiliation{Department of Physics, University of Connecticut, Storrs, Connecticut 06269, USA}
\affiliation{Department of Physics, Harvard University, Cambridge, Massachusetts 02138, USA}
\author{Susanne F. Yelin}
\affiliation{Department of Physics, Harvard University, Cambridge, Massachusetts 02138, USA}

\begin{abstract}
Finding a comprehensive and general description of the collective Lamb shift and cooperative broadening in a radiatively interacting system is a long-standing open question. Both energy levels and linewidth of individual atoms are modified by the exchange of real and virtual photons making up the dipole-dipole interaction. We introduce a method to theoretically study weakly-driven, low-excited ensembles of two-level atoms, and obtain an analytic description of the collective Lamb shift and linewidth via a self-consistent formalism including infinite order of correlations which stem from only two-body interactions. We predict the dependency of these quantities, as measurables, on system parameters: the number density of the ensemble, the detuning of an external probe field, and the geometry of the sample.
\end{abstract}


\maketitle


Ensembles of radiators can manifest collective effects as a consequence of dipole-dipole interactions mediated by photons. These effects can alter the radiative properties, including the buildup of collective modes called superradiance and subradiance  with enhanced or suppressed decay rates \cite{dicke1954,gross1982,guerin2016,araujo2016,das2020}, as well as the shift of transition frequency \cite{friedberg1973}. The collective Lamb shift \cite{lamb1947,scully2010}, which results from the exchange of virtual photons between radiators, has attracted much attention in both theoretical discussions and experimental studies \cite{friedberg2010,manassah2012,sutherland2016}. In the case of single excitation, the shift can be found by diagonalizing the equations of motion \cite{scully2009,svidzinsky2010,kong2017}. Meanwhile, classical electrodynamics simulations are also used to study spatially extended systems \cite{javanainen2014,javanainen2016,javanainen2017}. Experimental observations of collective Lamb shift have been reported in solid state samples \cite{rohlsberger2010}, atomic arrays \cite{meir2014} and atomic gas \cite{garrett1990,keaveney2012,bromley2016,jenkins2016,roof2016,jennewein2016,peyrot2018}. In most discussions to date, the studies of many-body radiative properties focus either on the buildup of collective modes \cite{asenjo2017array,patti2021,rubies2022}, or the effects of re-absorption and re-emission of real and virtual photons. Through collective modes, people have established a frequently employed pathway for studying super- and subradiance, while the re-absorption and re-emission of real and virtual photons are commonly utilized to study radiation trapping and vacuum effects. In this letter, we focus on the perspective of virtual photon exchange, in order to study the collective Lamb shift and the modification to linewidth in a dense sample. In contrast to the usual Lamb shift and natural linewidth arising from the interaction with the electromagnetic modes of the vacuum, in a dense sample the changed linewidth and shift results from the virtual interaction of electromagnetic modes with all the atoms. As the linewidth and the shift of a collective system are connected via the Kramers-Kronig relations, they need to be studied simultaneously.

A method often employed is the coupled-dipole model \cite{svidzinsky2010,araujo2018}, which explores superradiance and corresponding shift by solving the collective eigenmodes. In contrast, we want to study how the fundamental properties of individual atomic transitions, such as the energy levels and linewidths, are modified in the environment of many atoms with dipole-dipole interaction. For this, a more comprehensive method to describe virtual photon exchange is needed. A measurement protocol specific to collective Lamb shift should take into account the integrated effect of an ensemble, ideally as the effect on a probe field weak enough such that one can fully neglect excitations and light intensity which are at least second order in the Rabi frequency. As a comparison, many other treatments to date identify the collective Lamb shift as the frequency shift of transmitted light through a sample. However, the transmitted light measurement from an ensemble includes a combination of light-intensity-dependent effects (superradiance and corresponding frequency shift) and light-intensity-independent vacuum effects (spontaneous emission and Lamb shift). Our method is different from others as we manage to explicitly distinguish the two types of effects analytically, so the collective Lamb shift we describe purely stems from vacuum fluctuation and is irrelevant to the buildup of collective modes. Our pragmatic definition of collective Lamb shift is, therefore, the limiting case of collective shift with no excitation. The analytical framework of our method can be found in Ref.~\cite{ma2022}, where the light-intensity-dependent and independent effects are differentiated and then treated on an equal footing.

\begin{figure}
    \includegraphics[width=0.48\linewidth]{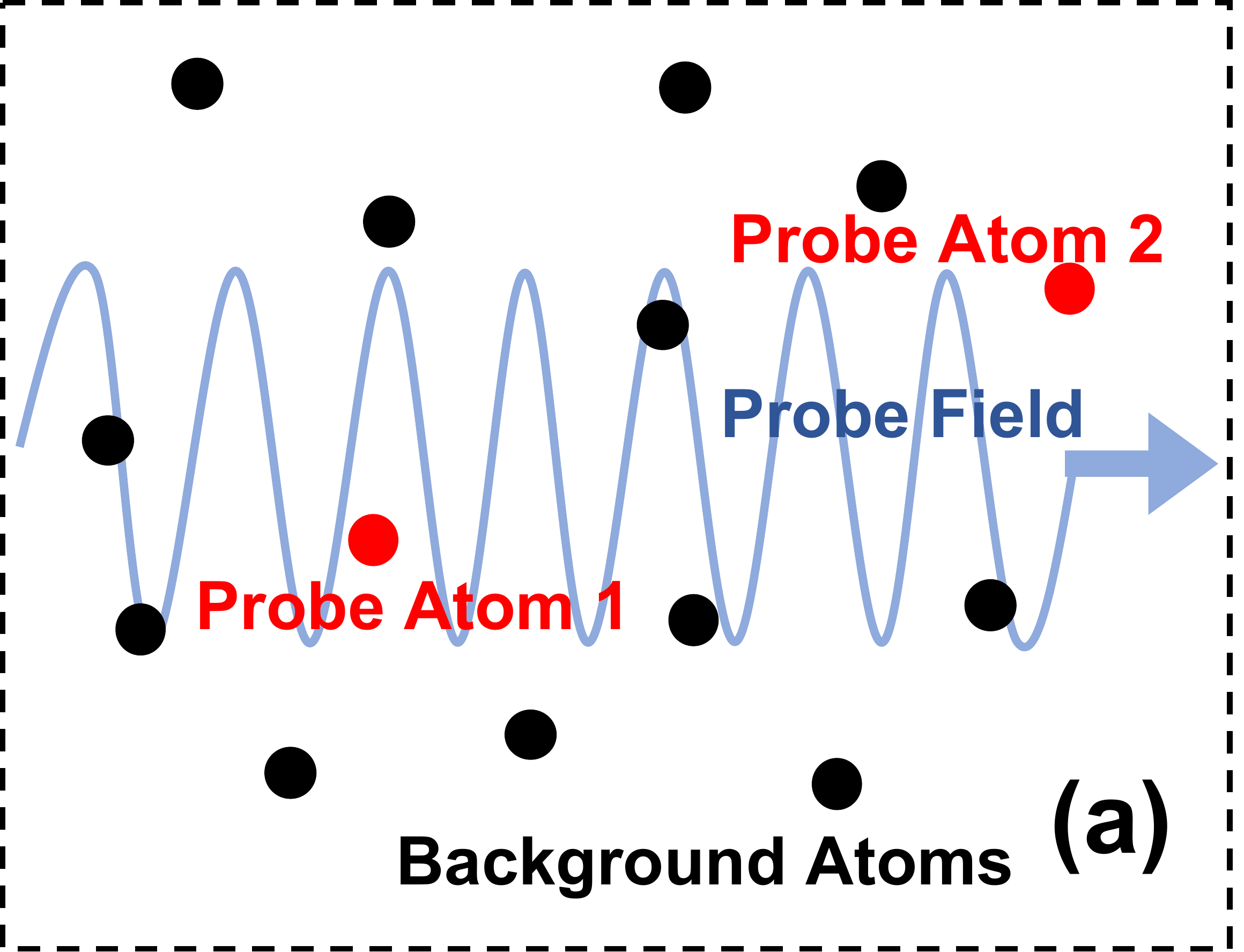}
    \includegraphics[width=0.48\linewidth]{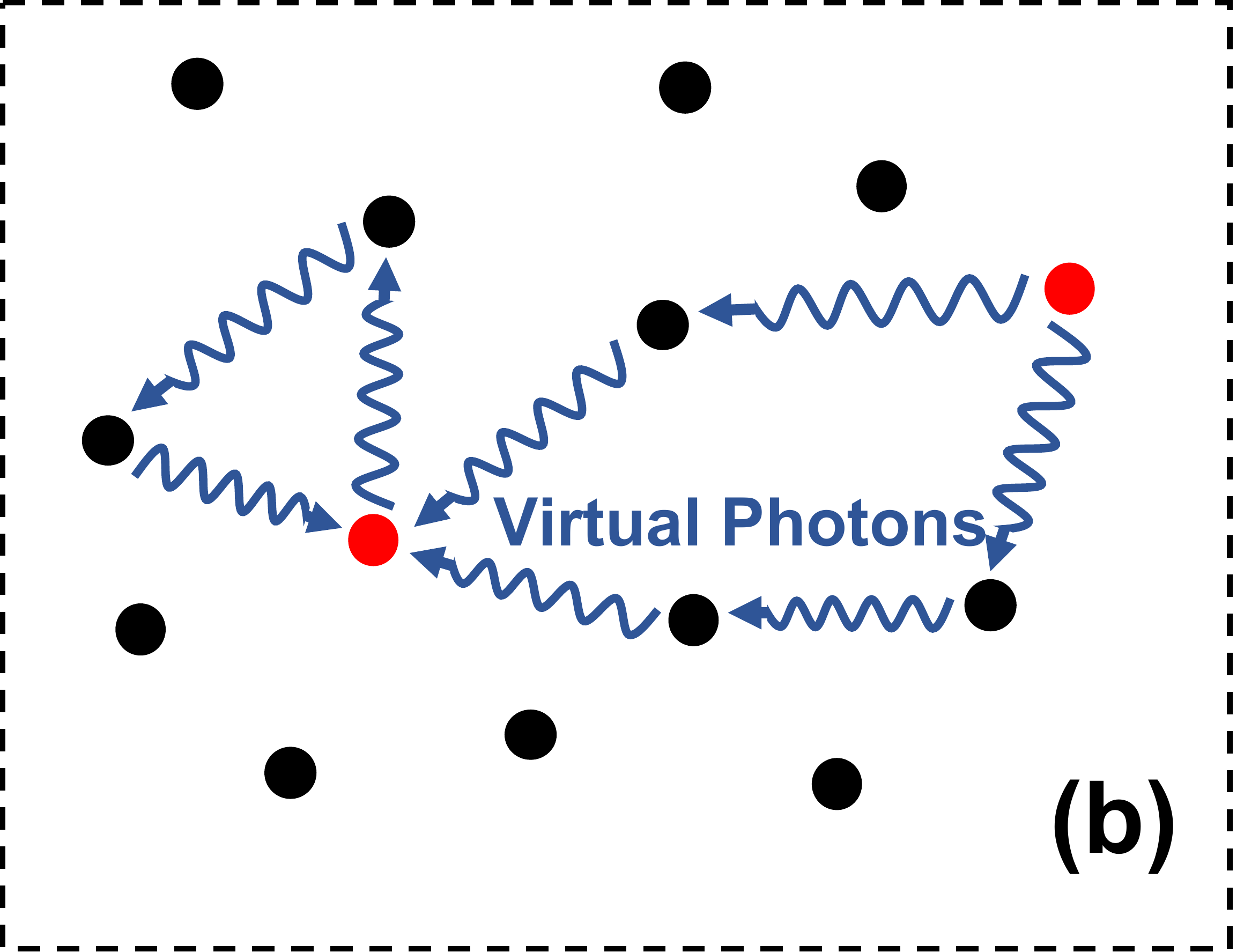}
    \caption{(a) A pictorial demonstration of the atomic gas driven by an external probe field. The reduced density matrix includes two arbitrarily chosen probe-atoms, while all other atoms as well as the field are traced out. (b) The emission and reabsorption of virtual photons result in the modifications of Lamb shift and spontaneous linewidth.}
    \label{fig:atomic_gas}
\end{figure}

In this Letter, we develop a self-consistent relation that analytically describes the collective Lamb shift and the linewidth of an interacting ensemble. This relation is governed by the local number density of atoms and the detuning of the external probe field. The scheme to derive such a relation can be summarized as follows: 
First, a master equation for any arbitrary choice of two probe atoms (\fig{fig:atomic_gas} (a)) is derived under the Markov approximation. The modified energy levels and linewidths of the probe atoms can be obtained by the real and imaginary parts of a dressed Green's function that describes the emission and re-absorption of real and virtual photons between probe and background atoms (\fig{fig:atomic_gas} (b)). Second, the dressed Green's function is calculated via the Dyson equation \cite{fetter}, and is expressed in terms of averaged two-point correlators of atomic operators. Third, by virtue of the quantum regression theorem \cite{meystre}, the averaged two-point atomic correlators are treated self-consistently and take into account multiple scattering of virtual photons to infinite order. This leads, finally, to a closed form of the modified linewidths and energy levels for a steady state. To predict measurable effects in experiments, one needs to take into account the geometry of the sample and average over the distribution of probe atoms. More details of the mathematical framework to this treatment can be found in Refs.~\cite{fleischhauer1999,ma2022}.

We consider an ensemble of identical two-level atoms that interact with the electric field, with the full Hamiltonian
\be\label{eq:hamiltonian}
    H=\sum_{i=1}^N \hbar\w_0 \s^\dagger_i \s_i+\sum_{\vec{k},\lambda}\hbar \w_l(\vec{k}) a^\dagger_{\vec{k},\lambda} a_{\vec{k},\lambda}-\sum_{i=1}^N \vec p_i\cdot (\vec {\cal E}_i+\vec E_i)
\ee
where $\s^\dagger_i=\ket{e_i}\!\!\bra{g_i}$ and $\s_i=\ket{g_i}\!\!\bra{e_i}$ are the raising and lowering operators of the $i$-th atom, $a^\dagger$ and $a$ are the creation and annihilation operators of the photons. We assume that the atoms mostly couple to the field modes with frequencies $\w_l$ that are close to the atomic resonant frequency $\w_0$. $\vec p_i=\hat{\epsilon}_i \wp(\s_i+\s_i^\dagger)$ is the dipole operator of the $i$-th atom with real dipole matrix element $\wp$. $\vec {\cal E}_i$ is the external classical driving field, while $\vec E_i$ is the quantized field operator, at the position of the $i$-th atom. Here, we focus on the dynamics of two arbitrarily chosen probe atoms, and formally trace out the quantized field and the remaining $N-2$ atoms \cite{meystre,lehmberg1970}. A Lindblad form master equation with rotating-wave approximation in the rotating frame is obtained for the two probe-atoms \cite{ma2022}
\beq\label{eq:master_eq}
\dot \rho &=& \ih \squ{\tilde{H}_0,\rho}+i\W \sum_{i=1,2} \squ{\sigma_{i} + \sigma_{i}^\dagger,\rho}-i\!\sum_{i,j=1,2}\!\delta_{ij}\squ{\s^\dagger_{j}\s_{i},\rho}\nonumber\\
&&-\sum_{i,j=1,2} \frac{\gamma_{ij}}{2} (\sigma_{j}^\dagger\sigma_{i}\rho - 2\sigma_{i}\rho\sigma_{j}^\dagger +\rho\sigma_{j}^\dagger\sigma_{i})
\eeq
\noindent where $\tilde{H}_0=\hbar(\w_0-\w_c)\sum_{i}\s^\dagger_{i}\s_{i}=\hbar\Delta_0\sum_{i}\s^\dagger_{i}\s_{i}$ is the two-atom free Hamiltonian in the rotating frame, with a classical driving frequency $\w_c$. $\W=|\wp {\cal E}_0/\hbar|$ 
is the Rabi frequency, with the driving amplitude ${\cal E}_0$. Note that a positive $\De_0$ stands for a red-detuned driving field. The two sets of parameters, $\g_{ij}$ and $\de_{ij}$, describe the modified spontaneous linewidth and the collective Lamb shift respectively.

It is commonly accepted that for atoms placed in free-space, $\g_{ij}$ and $\de_{ij}$ are associated with the free-space Green's tensor that results from the dipole-dipole interaction. However, this is no longer correct in a dense system, since one must consider the effect of the medium, namely, the multiple scattering of virtual photons between the particles (\fig{fig:atomic_gas} (b)). We define the dressed Green's function in the medium as
\be\label{eq:simple_greens_function}
D_{ij}(\tau,t)=D(\vec{r}_i,t+\tau,\vec{r}_j,t)= \theta(\tau)\sbrac{\squ{E^-_i(t+\tau),E^+_j(t)}}
\ee
and its Fourier transform with respect to $\tau$, $\tilde{D}_{ij}(\w,t)=\int_{-\infty}^{+\infty} D_{ij}(\tau,t)e^{-i\w\tau}d\tau$, where $E^\pm_{i(j)}(t)$ is the positive (negative) component of the field operator in the Heisenberg picture, and $\theta(\tau)$ is the Heaviside step function. The modified spontaneous linewidth and the collective Lamb shift are then written in terms of the dressed Green's function \cite{fleischhauer1999,ma2022}
\begin{subequations}\label{eq:gamma_ij_delta_ij}
\begin{align}
\g_{ij}(t) =& \frac{\wp^2}{\hbar^2}\int_{-\infty}^{+\infty} \sbrac{\squ{E_j^+(t),E_i^-(t+\tau)}}e^{-i(\w_0-\w_l)\tau}d\tau\label{eq:gamma_ij}\\
\de_{ij}(t) =& -\frac{i\wp^2}{2\hbar^2}\int_0^{+\infty} \sbrac{\squ{E_j^+(t+\tau),E_i^-(t)}}e^{i(\w_0-\w_l)\tau}d\tau\nonumber\\
&+\frac{i\wp^2}{2\hbar^2}\int_0^{+\infty} \sbrac{\squ{E_j^+(t),E_i^-(t+\tau)}}e^{-i(\w_0-\w_l)\tau}d\tau\label{eq:delta_ij}
\end{align}
\end{subequations}
thus $D_{ij}\!\sim\! \g_{ij}/2+i\de_{ij}$. These expressions can be formally derived by tracing out the $N-2$ background atoms and the quantized field \cite{supp}. For two arbitrarily chosen probe atoms, \eq{eq:master_eq} is effectively averaged over all possible choices of atomic pairs, thus a permutational symmetry is imposed such that $\gamma_{11}=\gamma_{22}$, $\gamma_{12}=\gamma_{21}$, and similarly for $\de_{ij}$. A Comprehensive derivation of Eqs. (\ref{eq:master_eq}) and (\ref{eq:gamma_ij_delta_ij}) for a dense atomic gas can be found in Ref. \cite{ma2022}, where the authors also take into account the contribution to the broadening and frequency shift which scales with the number of photons. Here, we consider a low-excited gas and neglect those effects. One can reproduce the free-space Green's tensor by using the free-field operators in \eq{eq:simple_greens_function} \cite{fleischhauer1999}, resulting in a function of distance between two spatial points \cite{gruner1996,asenjo2017array,patti2021,rubies2022}
\beq\label{eq:free_space_greens_function}
\tilde{D}^{(0)}_{\alpha\beta}(r,\w)&=&-\frac{i\hbar e^{-i\w r/c}}{4\pi\epsilon_0 r}\bigg[\delta_{\alpha\beta}\bigg(\frac{\w^2}{c^2}-i\frac{\w}{c}\frac{1}{r}-\frac{1}{r^2}\bigg)\nonumber\\
&&+\frac{x_\alpha x_\beta}{r^2}\bigg(-\frac{\w^2}{c^2}+3i\frac{\w}{c}\frac{1}{r}+\frac{3}{r^2}\bigg)\bigg]
\eeq
and the scalar free-space Green's function \footnote{If one looks at different polarizations, the Dyson equation here can be easily rewritten in the tensor form. But here we assume two-level atoms with transitions of only one polarization} by averaging over random polarization and replacing $x_\alpha x_\beta/r^2 \!\!\to\!\!\de_{\alpha\beta}/3$
\be\label{eq:free_space_scalar_greens_function}
\tilde{D}^{(0)}(r,\w)=-\frac{i\hbar \w^2}{6\pi \epsilon_0 c^2 r}e^{-i\w r/c}
\ee

In a dense sample, the dressed Green's function $D_{ij}$ obeys a general form of Dyson equation \cite{fetter,fleischhauer1999}
\beq\label{eq:dyson_eq}
D(\vec{r}_i,t_i,\vec{r}_j,t_j)=D^{(0)}(\vec{r}_i,t_i,\vec{r}_j,t_j)-\int\! dt_k\!\int\! dt_l\!\int\! d^3\! r_k \nonumber\\
\times D^{(0)}(\vec{r}_i,t_i,\vec{r}_k,t_k)P(\vec{r}_k,t_k,t_l)D(\vec{r}_k,t_l,\vec{r}_j,t_j)\nonumber\\
\eeq
which consists of the free-space term from \eq{eq:free_space_scalar_greens_function} as well as all orders of corrections that result from the multiple scattering of virtual photons.

The source function $P$ has the following form in a continuum approximation
\be\label{eq:source_function}
P(\vec{r}_i,t_1,t_2)=\frac{\wp^2 {\cal N}}{\hbar^2}\theta(t_1-t_2)\sbrac{[\s^\dagger_i(t_1),\s_i(t_2)]}
\ee
where ${\cal N}$ is the number density of the atoms. By iteration, the second term in \eq{eq:dyson_eq} includes a sum of all orders in $P$, describing multiple scattering of virtual photons to all orders. The correlation function on the right hand side of \eq{eq:source_function} can be further expressed in terms of the elements of the density matrix by using the quantum regression theorem \cite{supp}.

In Ref. \cite{fleischhauer1999,ma2022}, we solve \eq{eq:dyson_eq} in Fourier space for $\tilde{D}_{ij}$ for a gas of randomly polarized radiators:
\be\label{eq:greens_function}
\tilde{D}_{ij}(\w,t)=-\frac{i\hbar \w_0^2}{6\pi \epsilon_0 c^2 r}{\rm exp}\bigg(-i\frac{\w_0 r}{c}\sqrt{1+\frac{2i\hbar}{3\epsilon_0}\tilde{P}(\w,t)}\bigg)
\ee
where $r=|\vec{r}_i-\vec{r}_j|$ is the distance between the radiators. Here, the source function $\tilde{P}$ takes a spatial average and is independent of the location. Note that by setting $\tilde{P}=0$, \eq{eq:greens_function} reduces to \eq{eq:free_space_scalar_greens_function} in the randomly polarized condition, so the Green's function is indeed dressed by the source function $\tilde{P}$.

We further assume a weak-driving condition in the sample. Since the number of excitations scales as $O(\W^2)$ and we keep only up to $O(\W)$ terms, we will neglect all effects that depend on the number of excitations. This approximation is valid, for example, in systems described in Ref.~\cite{keaveney2012,peyrot2018}. We will approach the limit $\W\to 0$ in the expression of $\tilde{P}$. The Fourier variable in \eq{eq:greens_function} takes the value $\w=\w_0-\w_l\approx\de_{11}$. The source function is then simplified to \cite{supp}
\be\label{eq:P1ret}
\tilde{P}(\w\approx\de_{11},t)=\frac{\wp^2}{\hbar^2}{\cal N} \frac{-2}{\g_{11}-2i\De_0}
\ee
which depends on the density of atoms ($\cal N$), the modified single-atom linewidth ($\g_{11}$) and the detuning of the driving field ($\De_0$).

\noindent Combining \eq{eq:simple_greens_function} and \eqs{eq:gamma_ij_delta_ij}:
\begin{subequations}\label{eq:gamma_ij_delta_ij_greens function}
\begin{align}
\g_{ij}&=-2\frac{\wp^2}{\hbar^2} {\rm Re}\big(\tilde{D}_{ij}\big)\label{eq:gamma_ij_greens function}\\
\delta_{ij}&= \frac{\wp^2}{\hbar^2} {\rm Im}\big(\tilde{D}_{ij}\big)\label{eq:delta_ij_greens function}
\end{align}
\end{subequations}
For simplicity, we define the effective wave number $q_0$
\be\label{eq:q_0}
q_0=q_0'+i q_0''=\frac{\w_0}{c}\sqrt{1+\frac{2i\hbar}{3\epsilon_0}\tilde{P}}
\ee
By taking the limit $r\to 0$, the single-atom spontaneous linewidth is associated with the real part of the effective wave number $q_0$:
\be\label{eq:gamma_11}
\g_{ii}=\g_{ij}(r\to 0)=-2\frac{\wp^2}{\hbar^2}\lim_{r\to 0}
 {\rm Re}\big(\tilde{D}_{ij}\big)=\g_0\frac{\lambda}{2\pi}q_0'
\ee
where $\lambda=2\pi c/\w_0$ is the wavelength of the transition, and $\g_0=\wp^2 \w^3_0/3\pi \hbar \epsilon_0 c^3$ is the free-space spontaneous linewidth. For the Lamb shift, we renormalize to the free-space value:
\beq
\de_{ii}'&=&\de_{ij}(r\to 0)-\de_{ij}^{(0)}(r\to 0)\nonumber\\
&=&\frac{\wp^2}{\hbar^2} \lim_{r\to 0}{\rm Im}\big(\tilde{D}_{ij}-\tilde{D}^{(0)}_{ij}\big)=-\g_0\frac{\lambda}{4\pi}q_0''\label{eq:delta_11}
\eeq
By substituting \eq{eq:gamma_11} and \eq{eq:delta_11} to \eq{eq:q_0} and making use of the form in \eq{eq:P1ret}, one can obtain a self-consistent relation for the single-atom terms $\gamma_{ii}$ and $\delta_{ii}'$
\be\label{eq:main_result}
1+\frac{2 C}{2\frac{\De_0}{\g_0}+i\frac{\g_{ii}}{\g_0}}=\bigg(\frac{\g_{ii}}{\g_0 }-i\frac{2\delta_{ii}'}{\g_0}\bigg)^2
\ee
where we have defined the cooperativity parameter $C=\frac{\lambda^3 {\cal N}}{4\pi^2}$ which is proportional to the number density of atoms. \eq{eq:main_result} is the first main result of this paper. In an ensemble of interacting atoms, both the spontaneous linewidth $\g_{ii}$ and the Lamb shift $\delta_{ii}'$ of an individual atom are modified due to the collective nature of the dynamics. Under the weak-driving and low-excitation conditions, both quantities together satisfy \eq{eq:main_result} which is governed by the density of the radiators ($C$) and the detuning of the driving ($\De_0$).

\begin{figure}
    \includegraphics[width=0.48\linewidth]{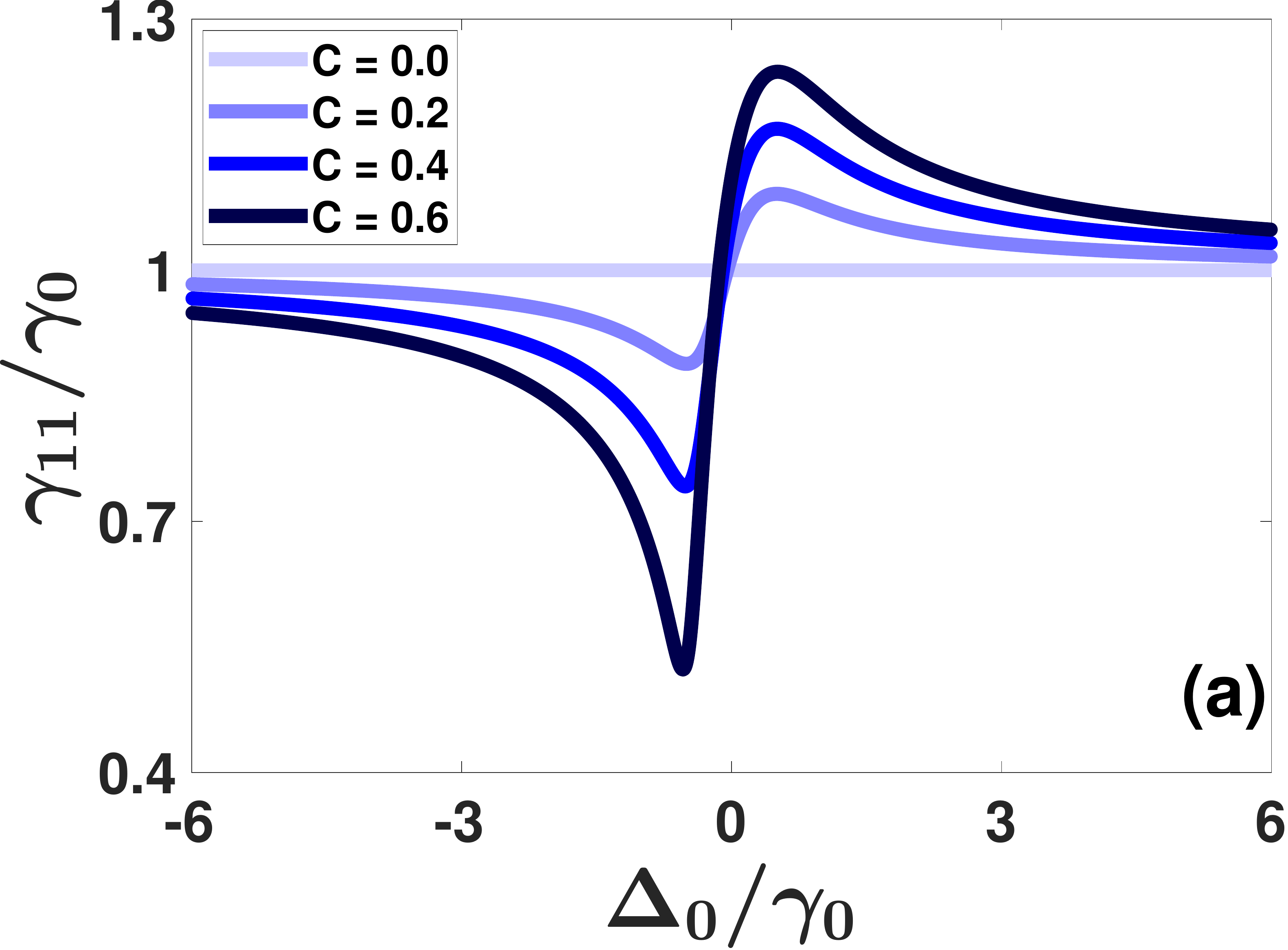}
    \includegraphics[width=0.48\linewidth]{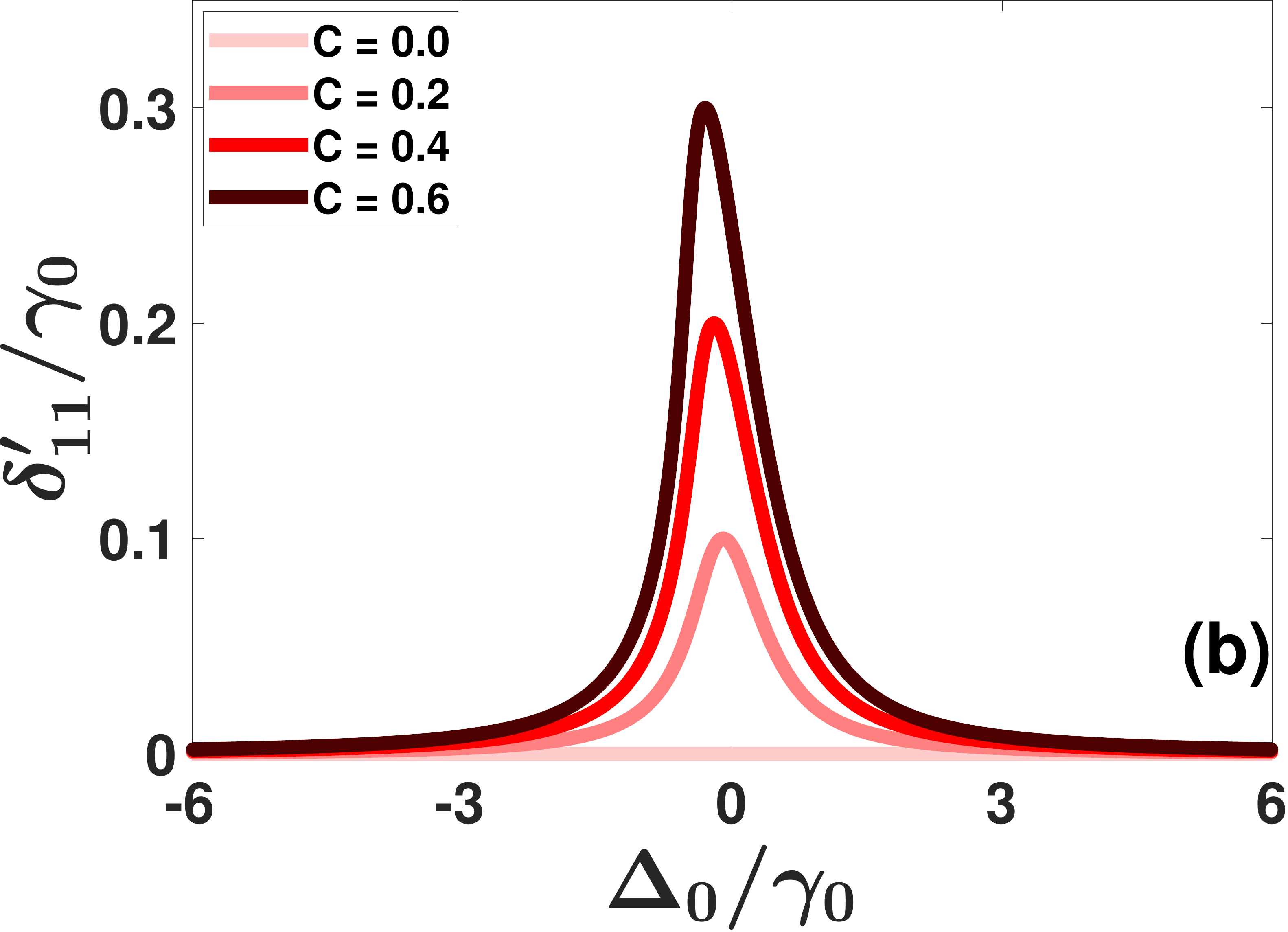}
    \includegraphics[width=0.48\linewidth]{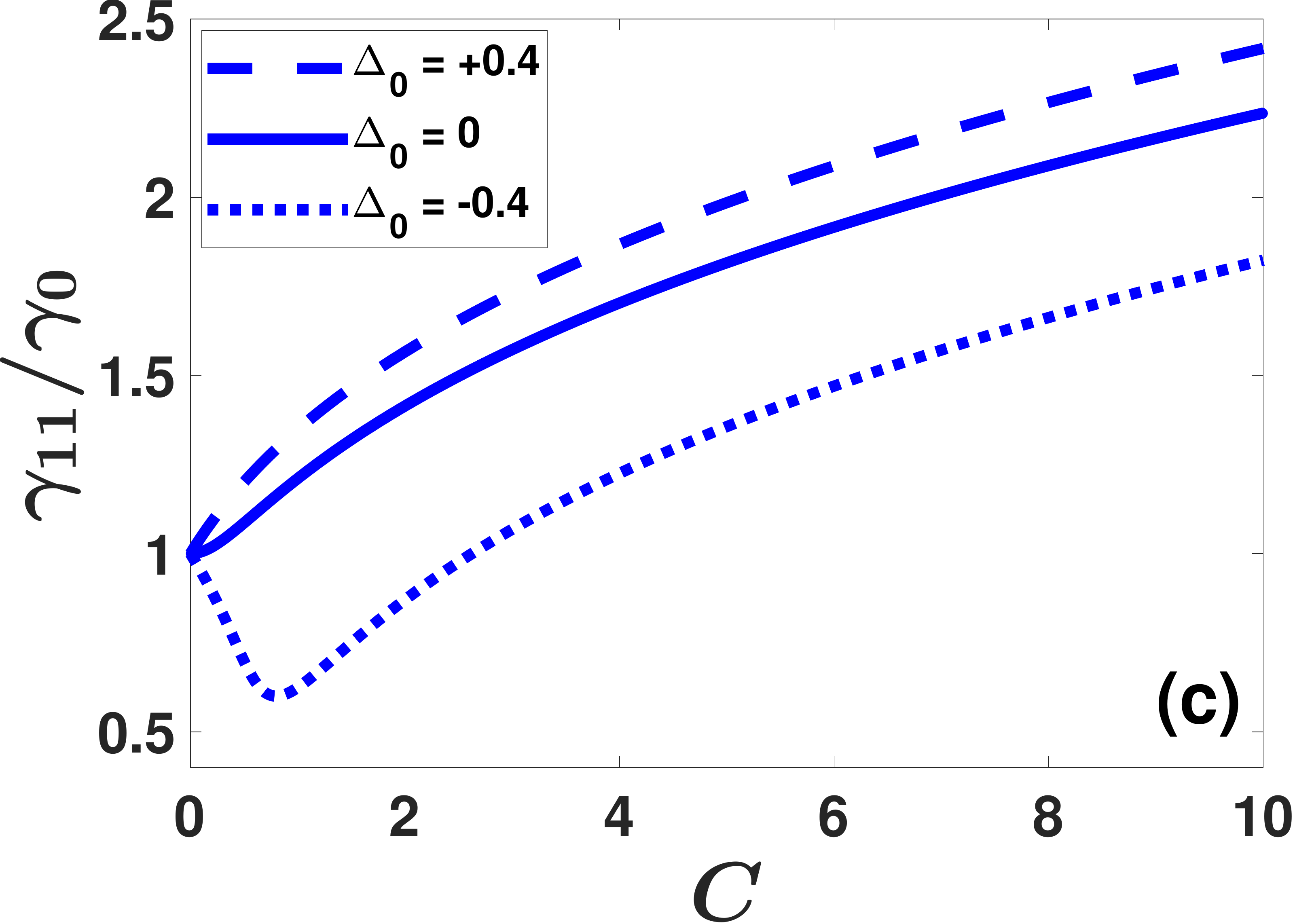}
    \includegraphics[width=0.48\linewidth]{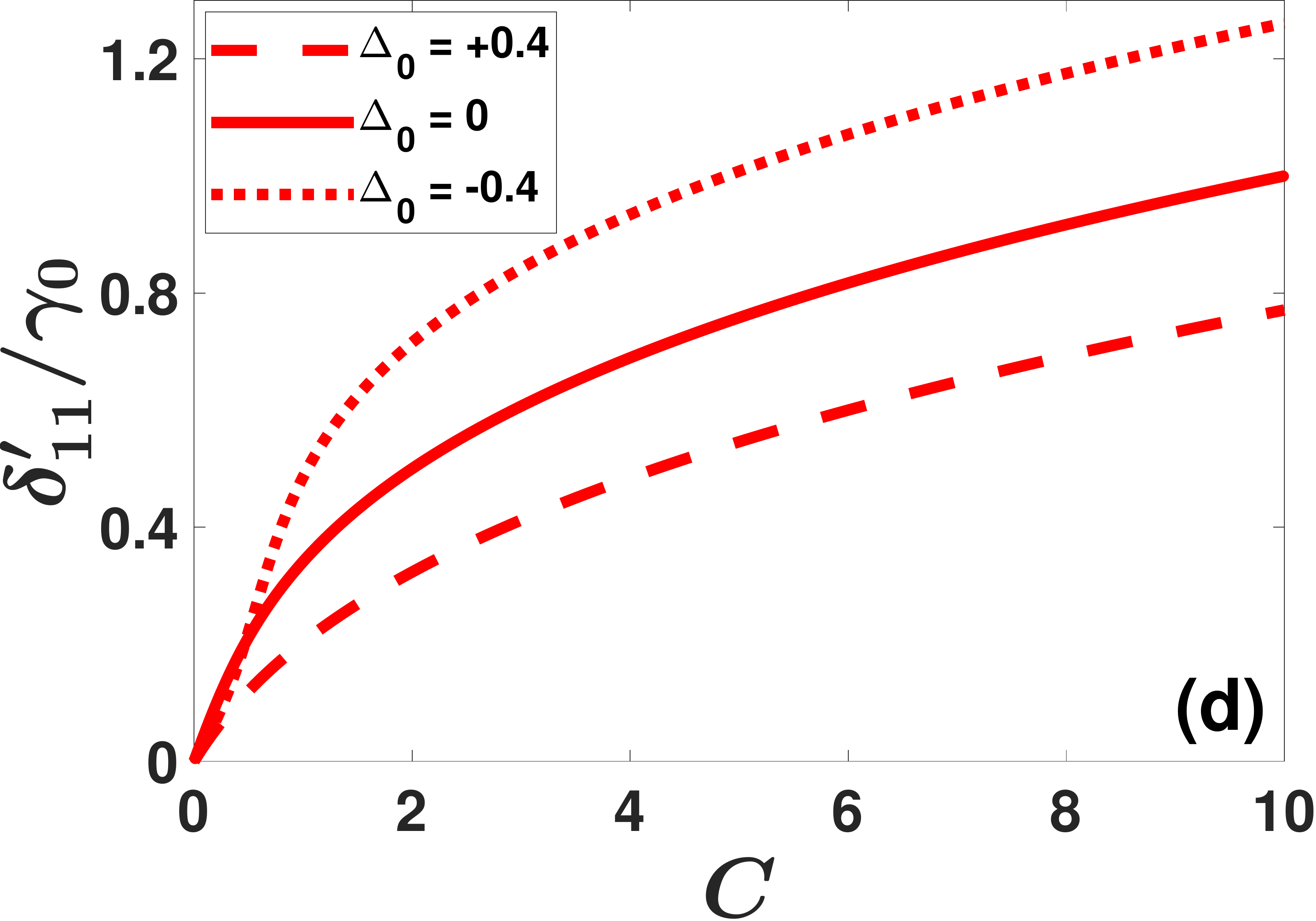}
    \caption{(a) The single-atom spontaneous linewidth modified by the collective effect, in units of the free-space linewidth $\g_0$. The value changes with respect to the detuning of the external driving field ($\De_0$), and the number density of emitters ($C$). (b) The single-atom Lamb shift modified by the collective effect, as a function of detuning. (c) The single-atom spontaneous linewidth as a function of particle density, with three different detunings. (d) The single-atom Lamb shift as a function of particle density.}
    \label{fig:gamma_11_delta_11}
\end{figure}

\fig{fig:gamma_11_delta_11} shows the solution of \eq{eq:main_result}, which has a modified Lorentzian profile. By varying the detuning of the external driving field and the density of particles, one can tune both the modified linewidth and the Lamb shift. At low-density or in far-detuned regions, both quantities recover the free-space values, namely, $\g_{11}\to\g_0$ and $\de_{11}'\to 0$. As the particle density gets higher, the collective effect predominates and leads to modification of both quantities. It is clear to see from \fig{fig:gamma_11_delta_11}(a) that a red-detuned driving field leads to an enhancement of spontaneous emission, while a blue-detuned driving field results in a suppression. A positive $\de_{11}'$ stands for a red shift of radiated light in our definition.

In \fig{fig:gamma_11_delta_11}(c,d) we plot the linewidth and the Lamb shift as functions of particle density, with different external detunings. Here, the atomic density $C=1$, for example, corresponds to $\sim\! 40$ particles contained in a cubed wavelength, which is $\sim\! 8\times 10^{13}/cm^3$ for Rb 780 nm transition. The minimum of $\g_{11}$ around $C=0.8$ corresponds to the valley at the blue-detuned region in \fig{fig:gamma_11_delta_11}(a). In the low density region ($C<0.5$), the Lamb shift increases linearly, while for high density, it exhibits sub-linear behavior.

\begin{figure}
    \includegraphics[width=0.48\linewidth]{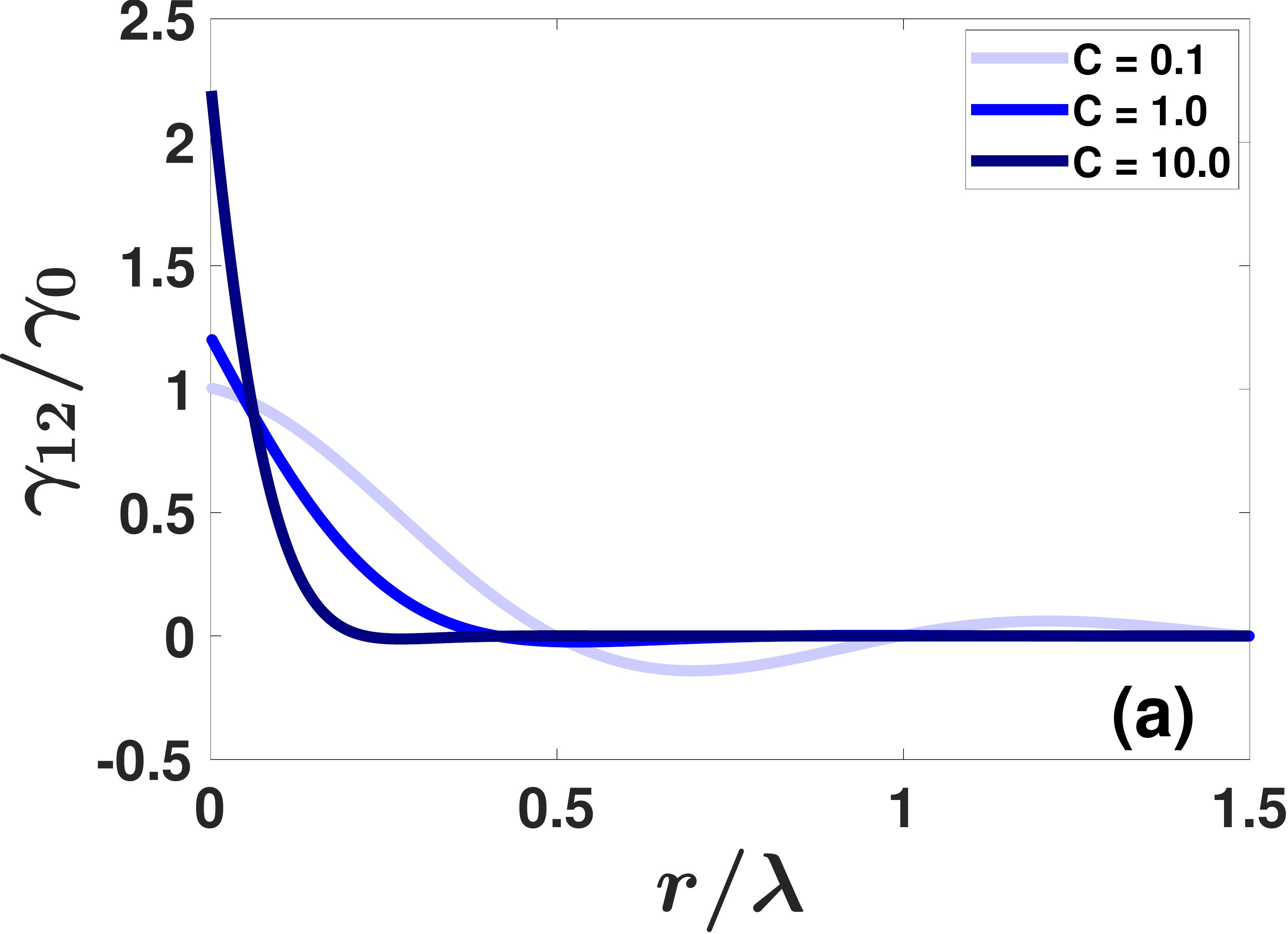}
    \includegraphics[width=0.48\linewidth]{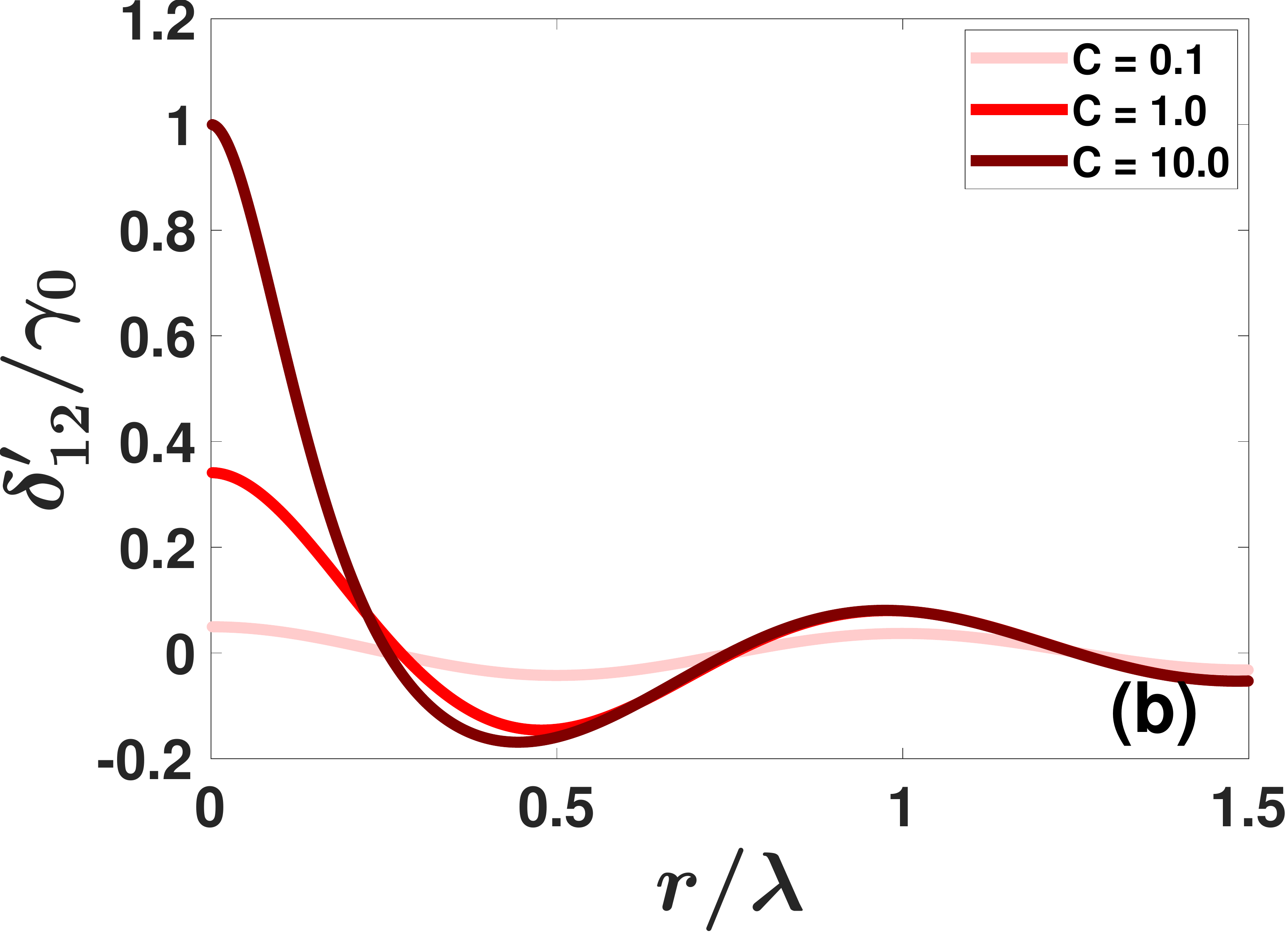}
    \caption{The two-atom terms of (a) modified linewidth and (b) Lamb shift as functions of distance between the two particles ($r$) in units of transition wavelength, with different particle density ($C$). In both calculations the driving is on-resonant with $\De_0=0$.}
    \label{fig:gamma_12_delta_12_change_r}
\end{figure}

The two-atom terms $\g_{12}$ and $\de_{12}$ can be calculated from \eq{eq:gamma_ij_delta_ij_greens function} as functions of inter-atom distance, and are shown in \fig{fig:gamma_12_delta_12_change_r}. An oscillatory behavior is shown, which results from the constructive/destructive contributions from the paired atom. In the small $r$ limit, both quantities reduce to the single-atom values, while in the large $r$ limit both vanish.

So far, we have been looking at the linewidth and energy shift for any single atom. But what are the measurable effects for the whole atomic gas? After choosing two probe atoms, both the dynamics and the steady states of the system can be solved via \eq{eq:master_eq}, which depend on the positions of the probe atoms. There is no obvious way to average over $\g_{ij}$ or $\de_{ij}$, but one can average over single-atom density matrices. Since the diagonal elements of the density matrices are trivially 1 and 0 for the ground and excited state populations in the weak excitation limit, the only important quantity is the off-diagonal element $\sbrac{e|\rho|g}$ which is related to the susceptibility. Starting from the ensemble average of the two-atom density matrix over all possible choices of probe atom pairs, $\rho^{\rm eff}=\frac{1}{n(n-1)}\sum_\alpha \rho^{(\alpha)}$ where $n$ is the total particle number and $\alpha$ denotes choices of probe atoms, the effective susceptibility of the whole ensemble is related to the averaged density matrix by
\beq\label{eq:chi}
\chi &\propto& {\cal N}\frac{\sbrac{e_1|\rho^{\rm eff}|g_1}+\sbrac{e_2|\rho^{\rm eff}|g_2}}{2}\nonumber\\
&=&\frac{1}{n(n-1)}\sum_\alpha {\cal N}\frac{\sbrac{e_1|\rho^{(\alpha)}|g_1}+\sbrac{e_2|\rho^{(\alpha)}|g_2}}{2}
\eeq
namely, one can first calculate the matrix elements $\sbrac{e_i|\rho^{(\alpha)}|g_i}$ for each probe atom pair, and then average over the ensemble.

For a weak driving, the $\W$ term in \eq{eq:master_eq} can be treated as a perturbation. \eq{eq:master_eq} then leads to the steady-state equations of the two-atom density matrix in the first order of $\W$, under a particular choice of probe atom pair $\alpha$:
\begin{subequations}\label{eq:steady_state_eqs}
\begin{align}
0\!=&-\bigg[\frac{\g_{11}}{2}+i(\de_{11}+\De_0)\bigg]\rho_{eg}\!+\!\big(\frac{\g_{12}}{2}+i\de_{12}\big)m_{eg}+i\W\\
0\!=&-\bigg[\frac{3}{2}\g_{11}\!+\!\g_{12}\!+\!i(\de_{11}\!+\!\De_0)\bigg]m_{eg}\nonumber\\
&-\!\big(\g_{11}\!+\!\frac{\g_{12}}{2}\!-\!i\de_{12}\big)\rho_{eg}-i\W
\end{align}
\end{subequations}
where $\rho_{eg}=\frac{1}{2}(\sbrac{\s_1}+\sbrac{\s_2})$ is the average single-atom coherence, and $m_{eg}$ is defined as $m_{eg}=\frac{1}{2}(\sbrac{\s_1\s_{z2}}+\sbrac{\s_{z1}\s_2})$, with $\s_i=\ket{g_i}\!\!\bra{e_i}$ and $\s_{zi}=\ket{e_i}\!\!\bra{e_i}-\ket{g_i}\!\!\bra{g_i}$. The solution to \eq{eq:steady_state_eqs} is
\begin{subequations}
\begin{align}
\rho_{eg}&=\frac{\W}{\de_{11}+\de_{12}+\De_0-i\frac{\g_{11}+\g_{12}}{2}}\label{eq:reg_solution}\\
m_{eg}&=-\rho_{eg}
\end{align}
\end{subequations}
By substituting \eq{eq:reg_solution} to \eq{eq:chi}, one can calculate the effective susceptibility of the atomic gas. Here, we employ a continuous approximation and take the average via an integration over the sample.
\beq\label{eq:ensemble_average}
\rho^{\rm eff}_{eg} &=& \int\!\!\!\!\int d^3 r_1 d^3 r_2 f(\vec{r}_1)f(\vec{r}_2)\nonumber\\
&&\times\frac{\W}{\de_{11}+\de_{12}(r)+\De_0-i\frac{\g_{11}+\g_{12}(r)}{2}}
\eeq
where $f(\vec{r})$ is the particle distribution function and $r=|\vec{r}_1-\vec{r}_2|$.

Furthermore, one can ``reverse-engineer'' the effective linewidth $\g^{\rm eff}$ and frequency shift $\de^{\rm eff}$ by the following relation:
\be
\rho^{\rm eff}_{eg} = \frac{\de^{\rm eff}+i\frac{\g^{\rm eff}}{2}}{(\de^{\rm eff})^2+(\frac{\g^{\rm eff}}{2})^2}\W
\ee
That is, after calculating the susceptibility via \eq{eq:ensemble_average} for a given atomic distribution $f(\vec{r})$, one can immediately separate its real and imaginary parts and obtain the effective linewidth and frequency shift behind the susceptibility. $\g^{\rm eff}$ and $\de^{\rm eff}$ are the quantities that can be directly measured in experiments.

\begin{figure}
    \includegraphics[width=0.48\linewidth]{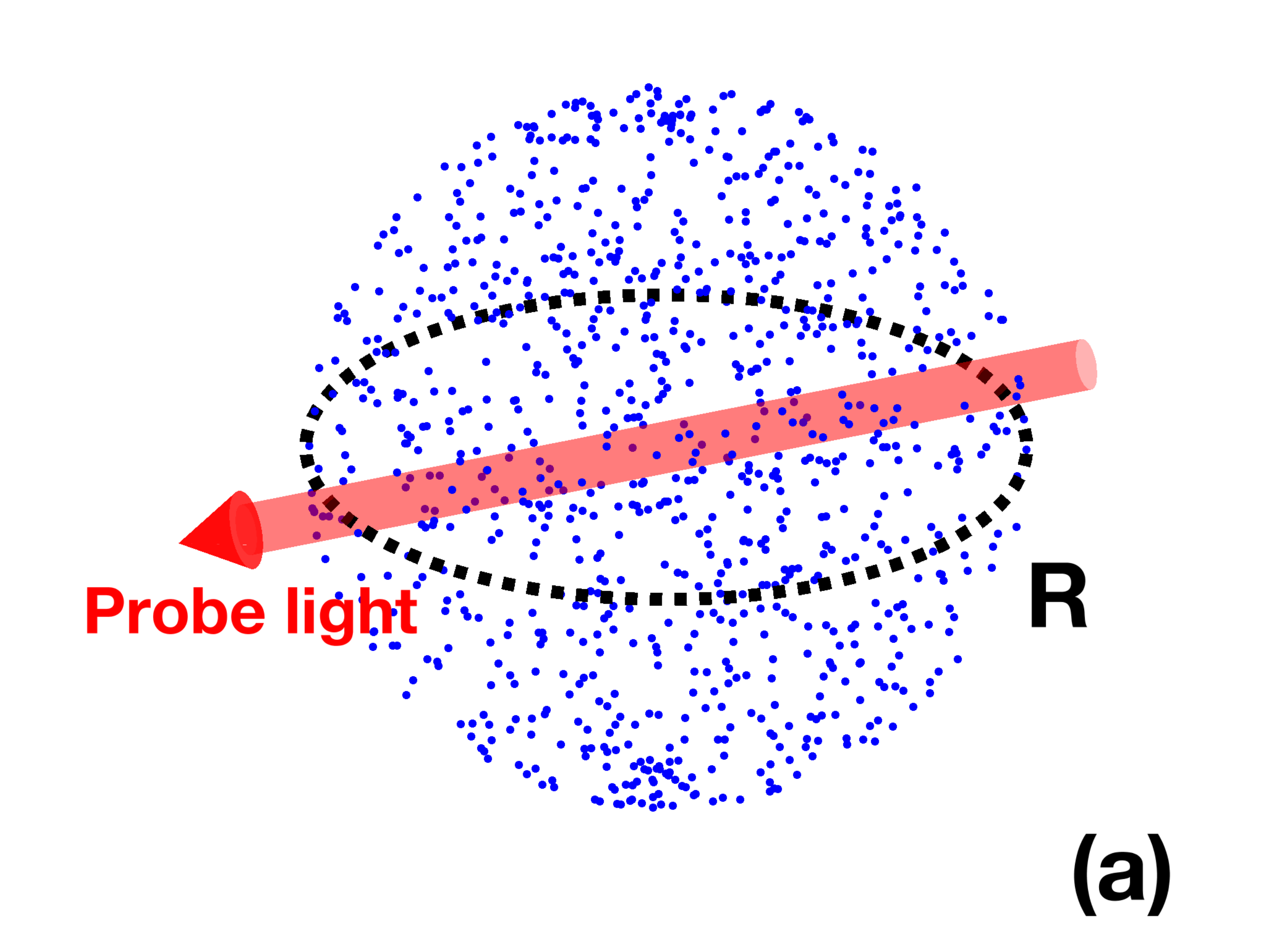}
    \includegraphics[width=0.48\linewidth]{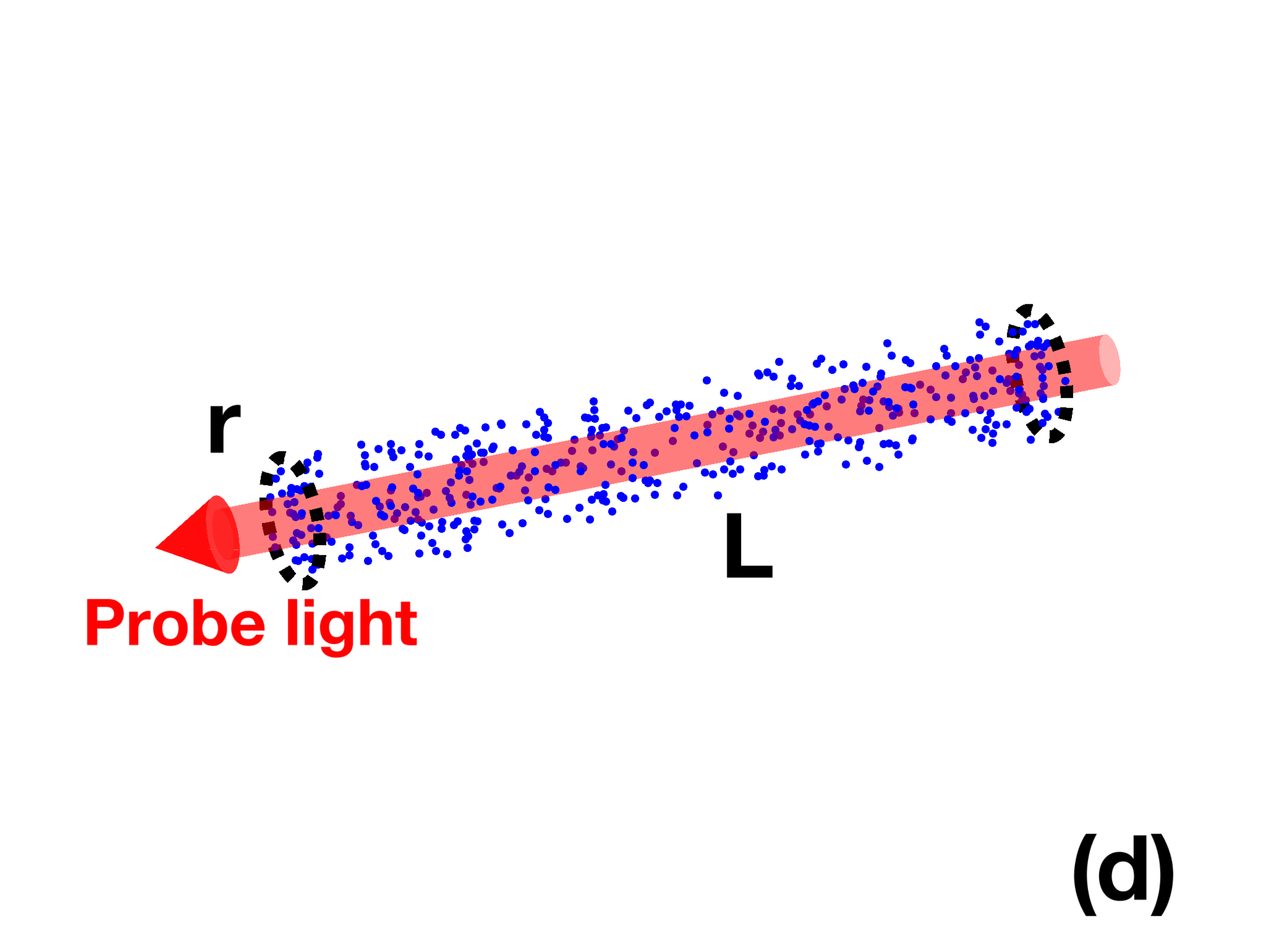}
    \includegraphics[width=0.48\linewidth]{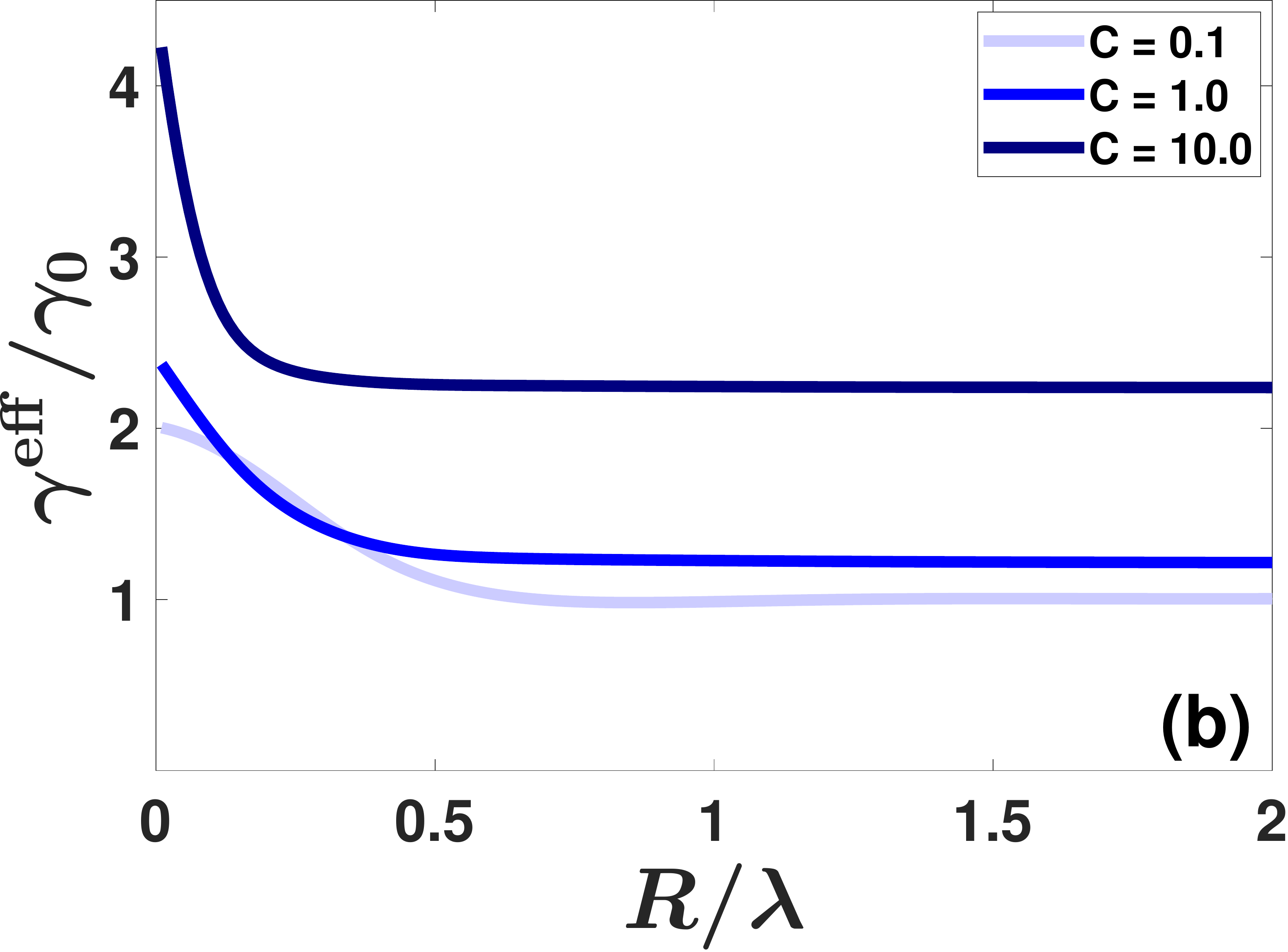}
    \includegraphics[width=0.48\linewidth]{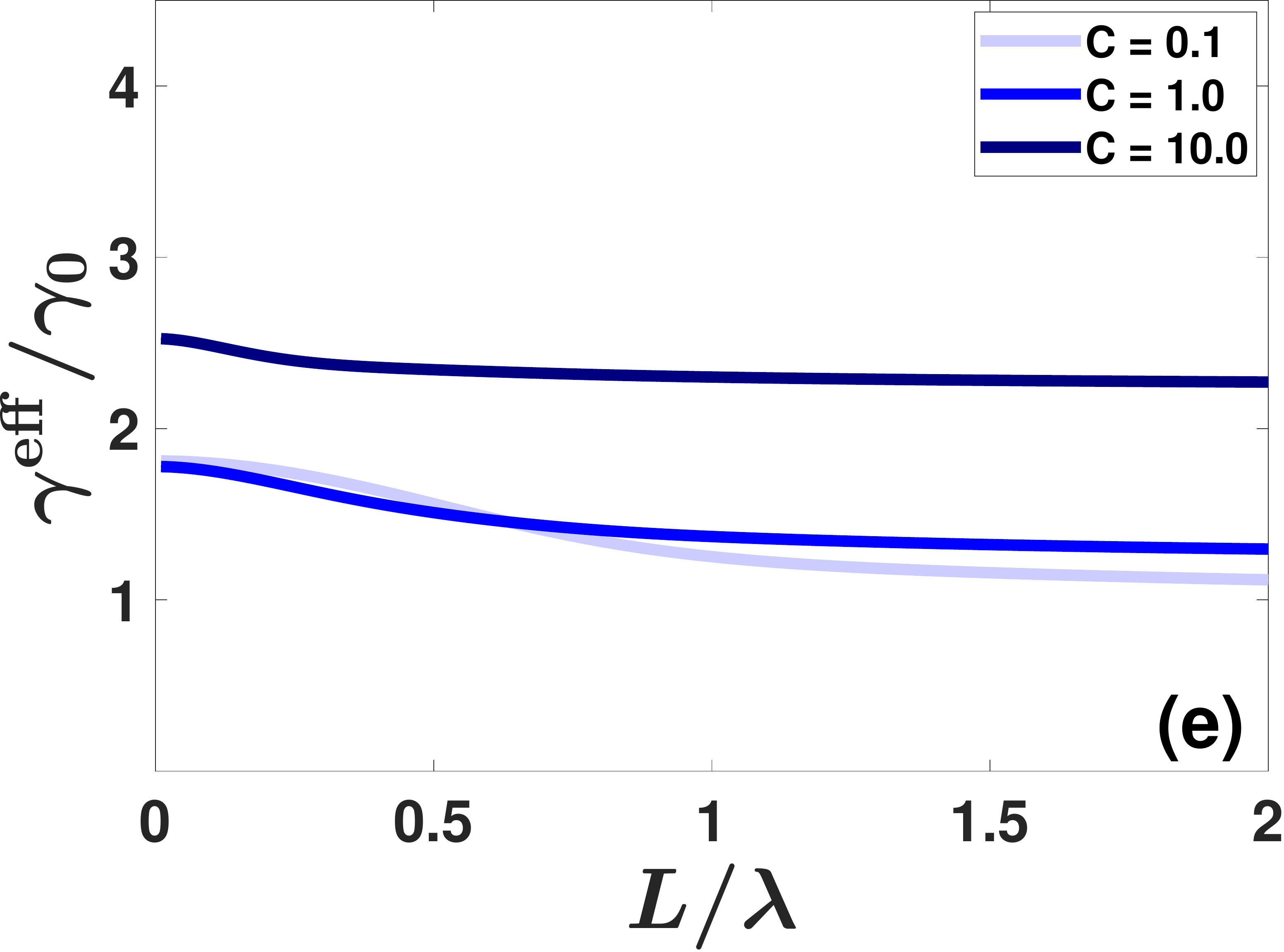}
    \includegraphics[width=0.48\linewidth]{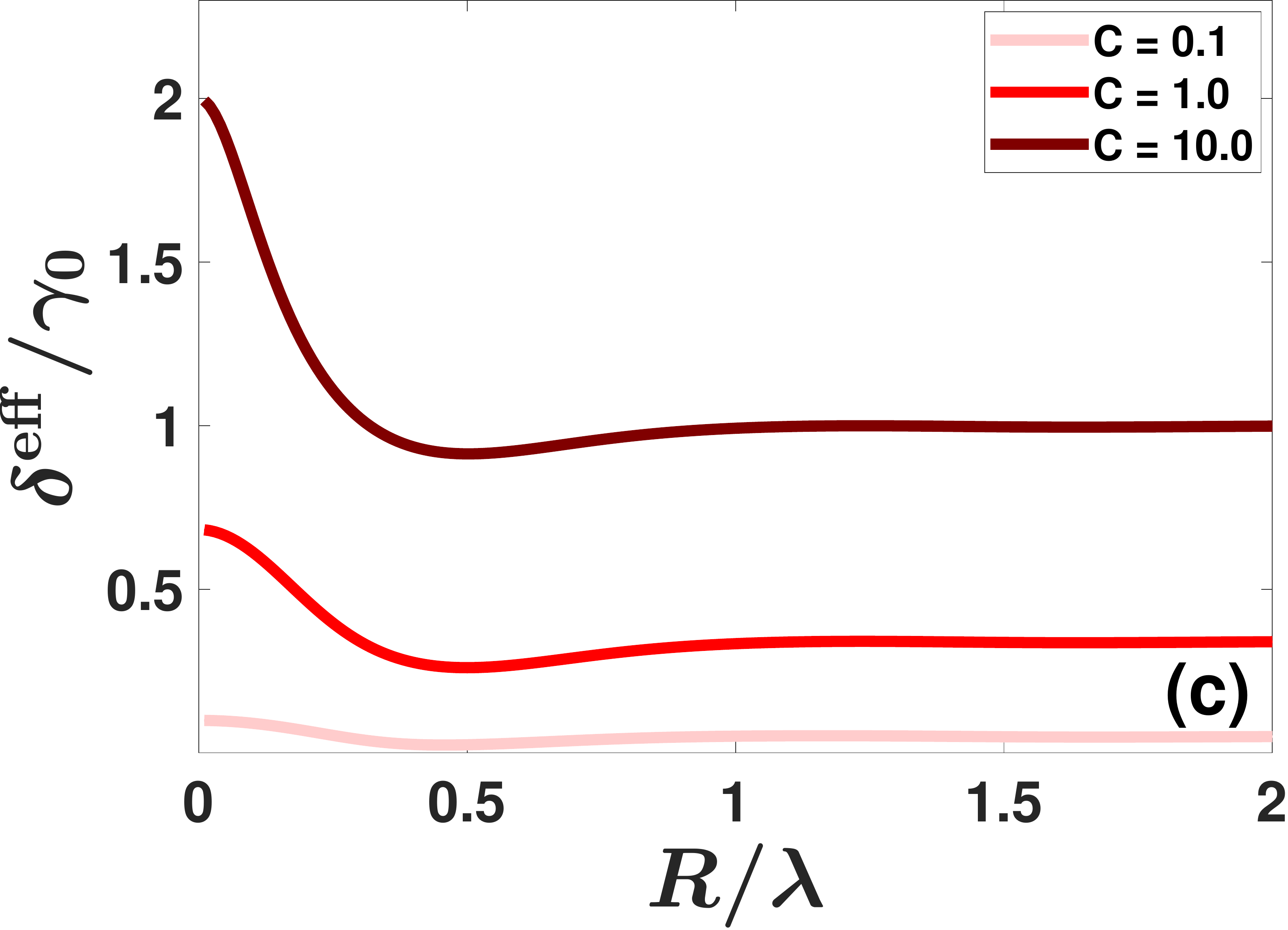}
    \includegraphics[width=0.48\linewidth]{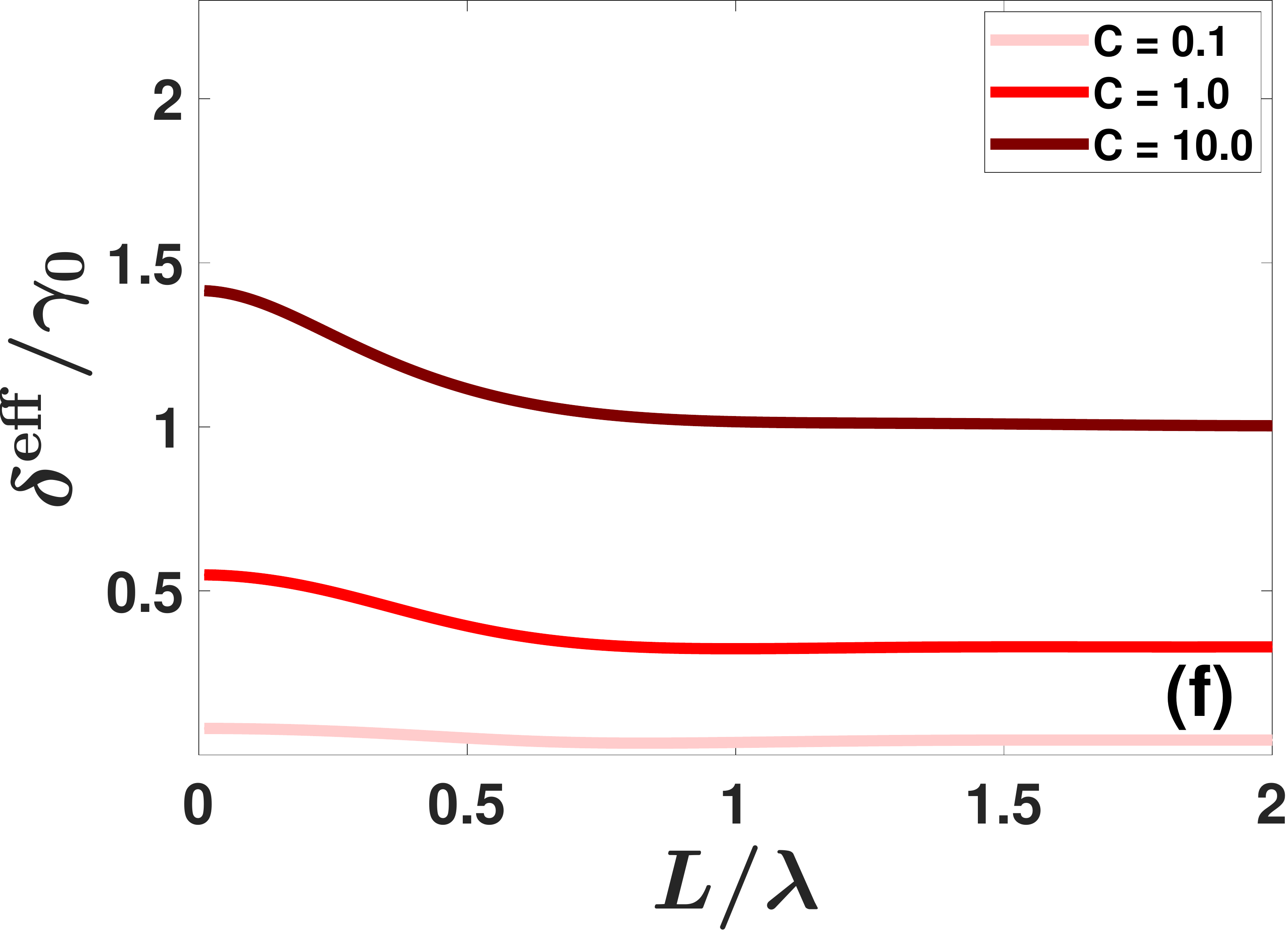}
    \caption{(a) A pictorial demonstration of a spherical gas of radiators with radius $R$. The propagation path of the probe light is drawn in red shadow. (b) The effective linewidth of a spherical gas, as a function of the radius, plotted with different particle densities. (c) The effective Lamb shift of a spherical cloud. (d) A pictorial demonstration of a cylindrical gas of radiators, with varying length $L$ and fixed radius $r=0.2\lambda$. (e) The effective linewidth of a cylindrical gas, as a function of the length, plotted with different particle densities. (f) The effective Lamb shift of a cylindrical gas.}
    \label{fig:sphere_cylinder}
\end{figure}

As an example, we consider a spherical gas of atoms with uniform distribution $f(\vec{r})=\frac{1}{V}$ and radius $R$ (\fig{fig:sphere_cylinder} (a)). Here, the first integration in \eq{eq:ensemble_average} is done over the whole sample, as each atom is influenced by every other atoms. The second integration is done along the propagation path of the probe light with a small waist compared to the transverse size of the sample, as these atoms along the path have major influence on the measured quantities. The result is shown in \fig{fig:sphere_cylinder} (b,c), where the effective, collective linewidth $\g^{\rm eff}$ and the collective Lamb shift $\de^{\rm eff}$ of the sample are plotted as functions of the sphere's radius, with different cooperativity parameter $C$. By and large, both $\gamma^{\rm eff}$ and $\delta^{\rm eff}$ decrease with larger radius of the atomic gas. As $R\to \infty$, both quantities asymptotically approach the values of the single-atom terms $\gamma_{11}$ and $\delta_{11}$, which are the homogeneous limits for an infinite size ensemble. With sufficiently high particle density, the collective Lamb shift shows an oscillatory behavior in small $R$. The minimum of $\de^{\rm eff}$ (least red shift) is found when the radius is near half of the transition wavelength. 

Similarly, a numerical result for a cylinder-shaped gas is shown in \fig{fig:sphere_cylinder} (e,f). This geometry resembles an elongated gas realized in recent experiments such as Ref. \cite{ferioli2023,glicenstein2021,ferioli2021}. The radius of the cylinder is fixed to $0.2\lambda$, and the modified linewidth and Lamb shift are plotted as functions of the length $L$. Both quantities also approach the homogeneous limits as the bulk volume increases. The oscillatory behavior as the bulk volume changes in both geometries is in qualitative agreement with some experiments \cite{keaveney2012,peyrot2018}, although they have slab-shaped gases instead. Exact matching in values is not achieved, because the transmitted light measured in experiments includes additional effects beyond vacuum fluctuation, such as the etalon effect from the parallel surfaces, which plays an important role in the measurement of light shift.

In conclusion, we have developed a theoretical framework to describe the collective radiation of dense ensembles. In particular, a weakly driven, low-excited condition is imposed, and the collective modification of Lamb shift and spontaneous linewidth are studied both analytically and numerically. We find that these quantities are connected via a dressed Green's function that describes the exchange and multiple scattering of virtual photons in the medium. The dressed Green's function, together with the master equation, lead to a self-consistent relation of linewidth and Lamb shift that shows their dependencies on the density of radiators and the detuning of the probe field. The linewidth and collective Lamb shift have a modified Lorentzian profile with respect to the detuning of the probe field, and the modification scales with the particle density. We find both enhancement and suppression of the decay rate, and an overall red-shift of the collective Lamb shift. We numerically predict the effective linewidth and Lamb shift that can be directly measured by experiments, and find oscillatory behavior as the size of the sample changes. These effects are at the order of the vacuum linewidth $\g_0$, and are already noticeable for a density low as $\sim\! 8\times 10^{13}/cm^3$ for Rb 780nm transition. Our work provides insights into the collective effects of radiative systems and is instructive to future experiments.

We thank Stefan Ostermann and Oriol Rubies-Bigorda for helpful discussions. We would like to thank the NSF for funding through PHY-1912607 and PHY-2207972 and the AFOSR through FA9550-19-1- 0233.

\bibliography{cls.bib}

\begin{widetext}
\section{Supplemental Material}
In this supplemental material, we provide the derivation of the two-atom master equation for a driven dense atomic gas that is used in the main text. The expressions of modified broadening terms $\gamma_{ij}$ and shift terms $\delta_{ij}$ are directly obtained from the master equation. To calculate the two-point correlators in the source function $P$, we make use of the quantum regression theorem and solve for the correlators in the Fourier space.

\subsection{I. Derivation of $\g_{ij}$ and $\de_{ij}$}

Here, We outline the derivation of the two-atom master equation and read off the expressions of $\g_{ij}$ and $\de_{ij}$. Details of the derivation is given in Ref. \cite{ma2022}. The full Hamiltonian can be separated into two parts
\beq
    H&=&\sum_{i=1}^N \hbar\w_0 \s^\dagger_i \s_i+\hbar \w_l a^\dagger a-\sum_{i\neq 1,2} \vec p_i\cdot (\vec {\cal E}_i+\vec E_i)-\sum_{i=1,2} \vec p_i\cdot (\vec {\cal E}_i+\vec E_i)\\
    &=& H_0 + V
\eeq
where $V=-\sum_{i=1,2} \vec p_i\cdot (\vec {\cal E}_i+\vec E_i)$ only contains two probe atoms. We implement the Keldysh formalism and define the effective two-atom time-evolution operator along the Keldysh contour
\be\label{eq:Seff}
\Se \equiv \sbrac{\SC}_E = \left\langle\TC {\rm exp}\left\{ -\frac{i}{\hbar}\int_{\cal C} d\check\tau V^I(\check\tau)\right\}\right\rangle_E
\ee
where $\sbrac{...}_E={\rm Tr}_E(...\rho_E)$ denotes an average over the environmental degrees of freedom, namely, the $N-2$ atoms and the quantized field. $V^I(t)=e^{iH_0 (t-t_0)/\hbar}V e^{-iH_0 (t-t_0)/\hbar}$ is the $V$ term in the interaction picture. $\cal C$ represents an integration path from $-\infty$ to $t$, then from $t$ to $-\infty$, with $\TC$ being the normal time-ordering for the first part and the inverse time-ordering for the second. The matrix element of the density matrix can be written as the expectation value of the corresponding projection operator.
\be\label{eq:rho_element}
\rho_{\alpha\beta}(t)= \sbrac{\TC \Se P^I_{\beta\alpha}(t)}_0
\ee
where $P^I_{\beta\alpha}(t)=e^{iH_0 (t-t_0)/\hbar}\ket{\beta}\!\!\bra{\alpha} e^{-iH_0 (t-t_0)/\hbar}$ and $\sbrac{...}_0={\rm Tr}(...\rho(t_0))$. Keeping the first and second order terms in \eq{eq:Seff} and taking the time derivative of \eq{eq:rho_element} gives rise to the two-atom master equation up to second-order correlations. The related terms in the master equation are
\beqq
\dot \rho &&= .....-\frac{\wp^2}{2\hbar^2}\!\!\sum_{i,j=1,2} (\sigma_{j}^\dagger\sigma_{i}\rho - 2\sigma_{i}\rho\sigma_{j}^\dagger +\rho\sigma_{j}^\dagger\sigma_{i})\int_{-\infty}^{+\infty} \sbrac{\squ{E_j^+(t),E_i^-(t+\tau)}}e^{-i(\w_0-\w_l)\tau}d\tau\\
&&+\frac{\wp^2}{2\hbar^2}\!\!\sum_{i,j=1,2}\squ{\s^\dagger_{j}\s_{i},\rho}\int_0^{+\infty} \!\!\bigg(\!\sbrac{\squ{E_j^+(t),E_i^-(t+\tau)}}e^{-i(\w_0-\w_l)\tau}\!-\!\sbrac{\squ{E_j^+(t+\tau),E_i^-(t)}}e^{i(\w_0-\w_l)\tau}\!\bigg)d\tau
\eeqq
The first term is the Lindblad term, with its coefficient defined as $\g_{ij}/2$, and the second term resembles the Hamiltonian time-evolution, with a coefficient defined as $-i\de_{ij}$. Therefore, we obtain the expressions for $\g_{ij}$ and $\de_{ij}$ in the main context.

\subsection{II. Derivation of the source function $P(\vec{r}_i,t+\tau,t)$ and its Fourier transform $\Tilde{P}(\vec{r}_i,\w,t)$}

In order to obtain the source function, one need to calculated the related two-point correlators of the atomic operators. To do this, we define the average excited population $\sbrac{\s_{ee}}$ and average single-atom coherence $\sbrac{\s}$ as
\beq
\sbrac{\s_{ee}}&=& (\sbrac{\s_{ee1}+\s_{ee2}})/2\\
\sbrac{\s}&=&(\sbrac{\s_1}+\sbrac{\s_2})/2
\eeq
where $\s_{eei}=\ket{e_i}\!\!\bra{e_{i}}$ and $\s_i=\ket{g_i}\!\!\bra{e_{i}}$. The master equation directly leads to the Maxwell-Bloch equations of motion
\beq
\frac{d}{d\tau}\sbrac{\s_{ee}} &=& -\g_{11} \sbrac{\s_{ee}}-i\W(\sbrac{\s}-\sbrac{\s^\dagger})\\
\frac{d}{d\tau}\sbrac{\s} &=& -\big[\g_{11}/2+i(\delta_{11}+\Delta_0)\big]\sbrac{\s} -i\W(2\sbrac{\s_{ee}}-1)\\
\frac{d}{d\tau}\sbrac{\s^\dagger} &=& -\big[\g_{11}/2-i(\delta_{11}+\Delta_0)\big]\sbrac{\s^\dagger} +i\W(2\sbrac{\s_{ee}}-1)
\eeq
The quantum regression theorem guarantees that two-time correlators follow the same equations of motion of single-operator expectation values \cite{meystre}:
\beqq
\frac{d}{d\tau}\sbrac{\s_{ee1}(t+\tau)\s_1(t)} &\!=\!& -\g_{11} \sbrac{\s_{ee1}(t+\tau)\s_1(t)}-i\W\big[\sbrac{\s_1 (t+\tau)\s_1(t)}-\sbrac{\s^\dagger_1 (t+\tau)\s_1(t)}\big]\\
\frac{d}{d\tau}\sbrac{\s_1 (t+\tau)\s_1(t)} &\!=\!& -\big[\g_{11}/2+i(\delta_{11}+\Delta_0)\big]\sbrac{\s_1 (t+\tau)\s_1(t)} -i\W(2\sbrac{\s_{ee1}(t+\tau)\s_1(t)}-\sbrac{\s_1(t)})\\
\frac{d}{d\tau}\sbrac{\s_1^\dagger(t+\tau)\s_1(t)} &\!=\!& -\big[\g_{11}/2-i(\delta_{11}+\Delta_0)\big]\sbrac{\s_1^\dagger(t+\tau)\s_1(t)} +i\W\big[2\sbrac{\s_{ee1}(t+\tau)\s_1(t)}-\sbrac{\s_1(t)}\big]
\eeqq
and similarly for $\sbrac{\s_1(t)\s_{ee1}(t+\tau)}$, $\sbrac{\s_1(t)\s_1 (t+\tau)}$ and $\sbrac{\s_1(t)\s_1^\dagger(t+\tau)}$. Note that $t$ is a fixed parameter and the initial conditions are such that $\tau=0$. Doing a Laplace transform with respect to $\tau$ helps to solve the differential equations easily, given initial conditions such as $\sbrac{\s_1^\dagger(t)\s_1(t)}=\sbrac{\s_{ee}(t)}$, $\sbrac{\s_1 (t)\s_1(t)}=0$ and $\sbrac{\s_1(t)\s_{ee1}(t)}=\sbrac{\s_1(t)}$. To obtain the Fourier transform of the two-point correlators, it suffices to replace the Laplace variable by $i\w$. Finally by substituting to Eq.~(8) in the main text, the Fourier transform of $P(\vec{r}_i,t+\tau,t)$ has the following form:
\beq
\tilde{P}(\vec{r}_i,\w,t)\!=\!\frac{\wp^2}{\hbar^2}{\cal N}\frac{(2\sbrac{\s_{ee}}\!-\!1)\big\{\big[\g_{11}/2\!+\!i(\de_{11}\!+\!\De_0\!+\!\w)\big](\g_{11}\!+\!i\w)\!+\!2\W^2\big\}\!+\!2\W\sbrac{\s}(-i\g_{11}/2\!+\!\de_{11}\!+\!\De_0\!+\!\w)}
{((\de_{11}+\De_0)^2+(\g_{11}/2+i\w)^2)(\g_{11}+i\w)+4\W^2(\g_{11}/2+i\w)}
\eeq
With the weak-driving and low-excitation approximations we set $\sbrac{\s_{ee}}=\sbrac{\s}=\W=0$, and by definition $\w=\w_0-\w_l\approx\de_{11}$. Expectation values are averaged over all possible probe atoms, and $\tilde{P}$ is independent of the locations at the end. It reduces to the following simple form
\be
\tilde{P}(\w\approx\de_{11},t)=\frac{\wp^2}{\hbar^2}{\cal N} \frac{-2}{\g_{11}-2i\De_0}
\ee
\end{widetext}

\end{document}